\documentclass{article}
\usepackage{spconf,amsmath,graphicx}
\usepackage{xcolor}
\usepackage{bm}
\usepackage{amsmath}
\usepackage{multirow}
\usepackage{booktabs}
\usepackage{diagbox}
\usepackage{tabularx}

\title{Utterance-level Neural Confidence Measure \\for End-to-End  Children Speech Recognition}
\name{Wei Liu, Tan Lee}
\address{ 
 DSP \& Speech  Technology  Laboratory \\
Department of Electronic Engineering, The Chinese University of Hong Kong \\
louislau\_1129@link.cuhk.edu.hk, tanlee@ee.cuhk.edu.hk}

\begin{document}
\topmargin=0mm 
%
\maketitle
\begin{abstract}
Confidence measure is a performance index of particular importance for automatic speech recognition (ASR) systems deployed in real-world scenarios. In the present study, utterance-level neural confidence measure (NCM) in end-to-end automatic speech recognition (E2E ASR) is investigated. The E2E system adopts the joint CTC-attention Transformer architecture. The prediction of NCM is formulated as a task of binary classification, i.e., accept/reject the input utterance, based on a set of predictor features acquired during the ASR decoding process. The investigation is focused on evaluating and comparing the efficacies of predictor features that are derived from different internal and external modules of the E2E system. Experiments are carried out on children speech, for which state-of-the-art ASR systems show less than satisfactory performance and robust confidence measure is particularly useful. It is noted that predictor features related to acoustic information of speech play a more important role in estimating confidence measure than those related to linguistic information. N-best score features show significantly better performance than single-best ones. It has also been shown that the metrics of EER and AUC are not appropriate to evaluate the NCM of a mismatched ASR with significant performance gap.
\end{abstract}

\section{Introduction}
Speech-enabled interactive systems, e.g., voice assistant, intelligent recorder, smart loudspeaker, etc., have been integrated widely into our daily life. Automatic Speech Recognition (ASR) is the most important technology underlying these systems. State-of-the-art ASR systems have been steadily improving in terms of recognition accuracy and processing latency. The system design has evolved from Hidden Markov Models (HMM) and hybrid model of Deep Neural Network and HMM (DNN-HMM) to attention-based end-to-end (E2E) neural model, which requires a substantially larger amount of labelled training data. Despite the continual efforts, the performance of an ASR system would be degraded inevitably in adverse and unstable acoustic conditions, e.g., high-intensity noise, atypical pronunciation and/or speaking style, and inadequately represented speaker groups. The present research is focused on children's speech.

Apart from the word error rate, confidence measure (CM) is a performance index of particular importance for ASR systems deployed in real-world scenarios. The value of confidence measure for an input speech utterance indicates to what level the user can trust the result of ASR. CM is useful in many downstream tasks of ASR. For example, it can be used to determine whether an input utterance (untranscribed) should be utilized for speaker adaptation \cite{uebel2001speaker}. CM is also useful in semi-supervised training \cite{chan2004improving}, active learning \cite{huang2016active}, spoken dialogue system \cite{hazen2002recognition,tur2005combining}, and intelligent audio stream allocation in distributed ASR systems \cite{kumar2020utterance}. 

Numerous approaches were investigated toward word-level CM in the context of HMM-based ASR \cite{jiang2005confidence}. Most commonly the CM is predicted with a binary classifier from features extracted during ASR decoding. The predictor features can be obtained from the ASR output lattice, for examples, word posterior probability \cite{evermann2000large}, word trellis stability \cite{sanchis2003estimating}, normalized acoustic likelihood and language model score. For the classifier models,  linear discriminant function \cite{sukkar1996vocabulary} , Gaussian mixture classifier \cite{chigier1992rejection}, decision tree \cite{neti1997word} and neural networks \cite{weintraub1997neural, kalgaonkar2015estimating,li2019bi, kastanos2020confidence} have been most commonly adopted.

An E2E ASR system is trained to realize sequence-to-sequence mapping typically via an encoder-decoder network \cite{chan2016listen,kim2017joint, dong2018speech}. Softmax probabilities in the auto-regressive decoder are commonly regarded as an intuitive measure of confidence on the mapping \cite{park2020improved}. However, the softmax probability was found to be unreliable and might perform poorly due to the overconfident behaviour of E2E models \cite{hendrycks2016baseline,li2020confidence}. To alleviate the problem of unreliability, a neural network can be trained independently to predict a softmax temperature value to re-distribute the original output probabilities at each time step of decoding \cite{woodward2020confidence}. In \cite{li2020confidence}, a lightweight neural network was used to estimate neural confidence measure (NCM), which was shown to be more reliable than directly using the softmax probability. In \cite{kumar2020utterance}, an NCM module was developed to predict utterance-level confidence measures in the context of small-footprint E2E ASR. With predictor features extracted from the encoder, the decoder and the attention blocks in the E2E system, the NCM significantly outperformed conventional word density confidence measure (WDCM) \cite{rueber1997obtaining} and beam-scatter weighted WDCM.

It is noted that the predictor features play a critical role in the design of robust NCM modules. In the present study we are focused to investigate predictor features that are discriminative to confidence measure and able to generalize well to other domains.
Here being discriminative means that the predictor features are effective in differentiating erroneous ASR outputs from correct ones.   
The feature of ``beam scores'' as investigated in \cite{kumar2020utterance} is extended. The efficacies of acoustic related and linguistic related score components are examined separately.


\begin{figure}[t]
    \centering
    \resizebox{6.7cm}{!}{\includegraphics{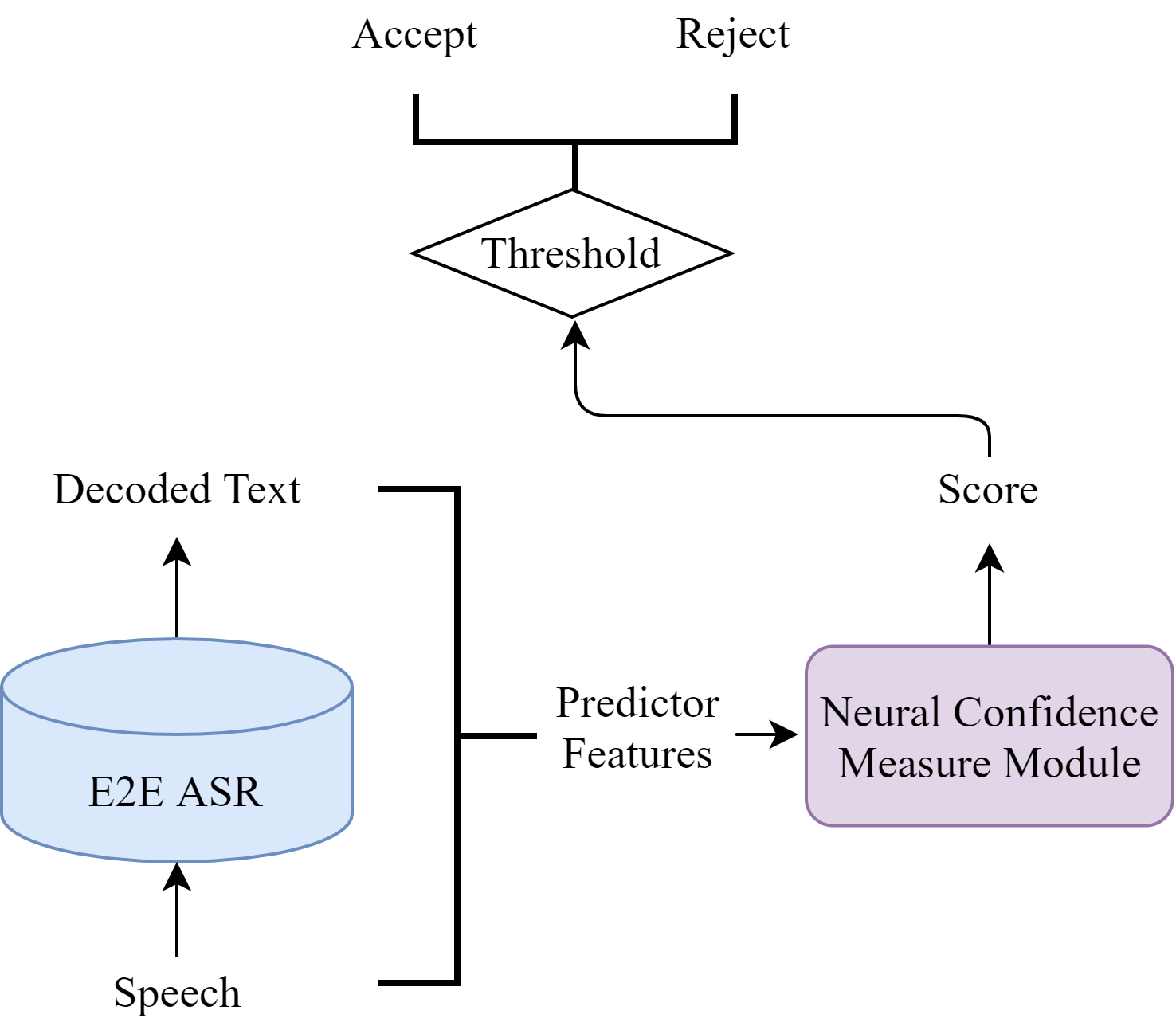}}
    \caption{Neural confidence measure prediction in E2E ASR}
    \label{fig:cm1}
\end{figure}

To our knowledge, this study is the first to explore utterance-level NCM in E2E ASR for children speech. Compared with adult speech, children speech is less studied and recently raises much research interests \cite{shivakumar2020transfer,shivakumar2021end}. ASR systems for children speech are more likely to generate the erroneous output, making confidence measure a more relevant issue than for adult speech. We also investigate the robustness of NCM to varying input speech conditions and the transferrability to another out-of-domain E2E adult ASR. In this paper, E2E ASR refers specifically to the \textit{Transformer} based joint CTC-attention speech recognition system. 

\section{Neural Confidence Measure Module}
As shown in Figure \ref{fig:cm1}, the NCM module is built on top of a properly trained E2E ASR system. For each input utterance decoded by the ASR system, the NCM module generates a confidence score based on a set of predictor features acquired from the ASR decoding process. The confidence score is compared against a threshold to determine whether the decoded text should be accepted or rejected.

\begin{figure}[t]
    \centering
    \resizebox{5.8cm}{!}{\includegraphics{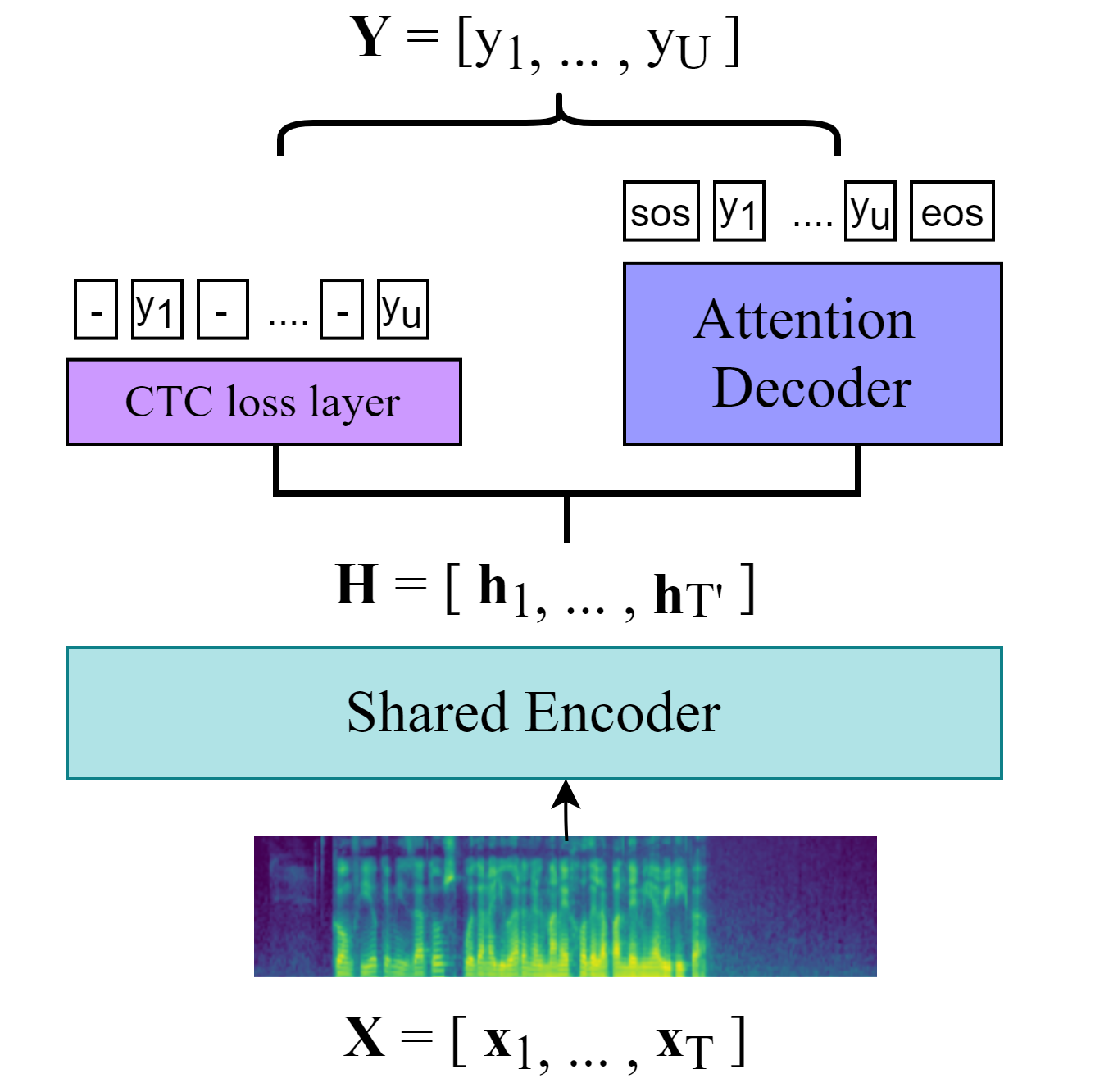}}
    \caption{Diagram of E2E speech recognition system}
    \label{fig:e2e_asr}
\end{figure}

\subsection{End-to-end speech recognition system}
In this study, the E2E ASR system is based on an encoder-decoder model with the joint CTC-attention learning framework. As shown in Figure \ref{fig:e2e_asr},  the system consists of three components: a shared encoder, an attention decoder and a connectionist temporal classification (CTC) loss layer. An input sequence of acoustic features, denoted as $\mathbf{X}= [\mathbf{x}_1,...,\mathbf{x}_T]$, is encoded by the shared encoder into a hidden vector sequence $\mathbf{H}=[\mathbf{h}_1, ...,\mathbf{h}_{T'}]$. The hidden sequence is then processed in parallel by the attention decoder and the CTC loss layer to generate a sequence of output tokens $\mathbf{Y}=[y_1,...,y_U]$. The three components are jointly optimized in training.



To take full advantages of both CTC and attention mechanisms, a multi-task learning (MTL) based loss function is used \cite{kim2017joint},
\begin{equation}
     \mathcal{L}_{MTL} = \lambda \mathcal{L}_{CTC} + (1-\lambda) \mathcal{L}_{Attention},
\end{equation}
where $\mathcal{L}_{CTC}$ denotes the CTC loss and $\mathcal{L}_{Attention}$ denotes the attention loss. The value of $\lambda$ is between $0.0$ and $1.0$. The CTC objective function acts as an auxiliary task to help speed up the alignment process at both training and decoding stages. The attention decoder relieves the limitation of conditional independence assumed in CTC.

\subsection{Predictor features}
\label{ssec:predictor_features}

Different network structures like feed forward network (FFN) \cite{li2020confidence,woodward2020confidence}, recurrent neural network (RNN) \cite{kalgaonkar2015estimating,kastanos2020confidence} and self-attention \textit{Transformer} \cite{kumar2020utterance} could be applied to realize the NCM module. In this study, a residual FFN with three hidden layers is adopted as the classification model. Our main focus is on selecting better predictor features. 

Design of predictor features depends highly on the ASR decoding process. In our E2E ASR system, the beam search algorithm is adopted to perform one-pass decoding \cite{watanabe2017hybrid}, in which the decoder computes a score for each partial hypothesis. In practice, external language models, e.g., n-gram and RNN language model (rnnlm), are employed by shallow fusion in the decoding. The score given to a recognition hypothesis would be the weighted combination of four components, namely the CTC score $\alpha_{ctc}$, the attention score $\alpha_{att}$, the n-gram score $\alpha_{ngram}$ and the rnnlm score $\alpha_{rnnlm}$.  They can be explained as the log probability to form the decoded token sequence from different information sources. The one with a higher score is more likely to be selected.
Here, the term ``token" refers to the decoded character at each time step. Each hypothesis is made up of a sequence of tokens.
 To speed up the search, only a limited number of partial hypotheses are retained at each time step according to the setting of $beam\_size$. An $N$-best list is generated at the end of decoding. The list includes $N$ complete hypotheses with the highest scores. The best hypothesis $\mathbf{\hat{Y}}_{best}$ is the final output of ASR, i.e.,
\begin{multline}
\label{eqn:eq4}
        \mathbf{\hat{Y}}_{best} = \operatorname*{arg\,max}_{\mathbf{\hat{Y}} \in \Omega} \{ \lambda_1 \alpha_{ctc}(\mathbf{\hat{Y}}|\mathbf{X}) + \lambda_2 \alpha_{att}(\mathbf{\hat{Y}}|\mathbf{X}) \\ + \lambda_3 \alpha_{ngram}(\mathbf{\hat{Y}}) + \lambda_4
    \alpha_{rnnlm}(\mathbf{\hat{Y}}) \},
\end{multline}
where $\Omega$ denotes a set of complete hypotheses, $\lambda_i$ are the component weights in the range of $[0, 1]$, satisfying $\lambda_1 +\lambda_2 =1$ and  $\lambda_3 +\lambda_4 =1$. Specifically, $\alpha_{ctc}(\mathbf{\hat{Y}}|\mathbf{X})$ and $\alpha_{att}(\mathbf{\hat{Y}}|\mathbf{X})$ refer to the negative CTC loss and attention loss respectively.

Intuitively, all of the four component scores are potentially contributive to estimating the confidence score. Note that the normalized score will be used to avoid the influence of various utterance length. Generally speech recognition models tend to assign extreme score value for the best hypothesis, which may adversely influence the classification judgment.
The top-$n$ best hypotheses' likelihoods could be utilized to produce a more robust confidence score. In the best hypothesis, apart from a list of utterance-level scores, more information from score distribution over the whole vocabulary can be utilized at each decoder time step. Due to the very large ASR vocabulary size and $beam\_size$ limitation in the beam search decoding, only top-$K$ score logits are kept to compute the softmax probability distribution and the average token entropy. In previous works \cite{kumar2020utterance, woodward2020confidence}, internal neural features of E2E ASR were studied. Embedding features extracted at the last layers of the encoder and the decoder were shown useful for confidence score estimation, given that they contain rich acoustic and linguistic information respectively. Token duration feature was also suggested based on the observation that tokens with short duration are prone to recognition errors. Table \ref{tab:pred_feat} lists the above predictor features, which are investigated in the following experiments.

\begin{table}[t!]
\centering
\caption{Basic predictor features derived from E2E ASR decoding process for NCM.}
\begin{tabular}{c|c|c}
\toprule
Predictor feature      & Feature form                                                                & Notation         \\ \midrule \midrule
CTC score              & \multirow{6}{*}{scalar}                                                     & $\alpha_{ctc}$   \\ \cline{1-1} \cline{3-3} 
attention score        &                                                                             & $\alpha_{att}$   \\ \cline{1-1} \cline{3-3} 
ngram score            &                                                                             & $\alpha_{ngram}$ \\ \cline{1-1} \cline{3-3} 
rnnlm score            &                                                                             & $\alpha_{rnnlm}$ \\ \cline{1-1} \cline{3-3} 
average token duration &                                                                             & $avg\_dur$       \\ \cline{1-1} \cline{3-3} 
average token entropy  &                                                                             & $ent$            \\ \hline
$top$-$n$ best scores       & vector                                                                      & $nbest\_score$    \\ \hline
encoder embedding      & \multirow{3}{*}{\begin{tabular}[c]{@{}c@{}}vector \\ sequence\end{tabular}} & $encfeat$        \\ \cline{1-1} \cline{3-3} 
decoder embedding      &                                                                             & $decfeat$        \\ \cline{1-1} \cline{3-3} 
$top$-$K$ score logits   &                                                                             & $nvocab\_scores$ \\ \bottomrule
\end{tabular}
\label{tab:pred_feat}
\end{table}

\section{Experimental Setup}
\subsection{Data sets}
The dataset used in the experiments on NCM are from the \textbf{2021 SLT Children Speech Recognition Challenge} (CSRC) \cite{yu2021slt}. The CSRC provided both adult and children speech data, each part being divided into training, validation (denoted as \textit{dev}) and test (denoted as \textit{test}) \cite{ng2020cuhk}. Additional evaluation sets released by the CSRC, namely the children read speech (\textit{eval\_child\_read}) and children conversational speech (\textit{eval\_child\_convers}) are also used in this study.
The NCM module is trained on the \textit{dev} part of children speech in the CSRC dataset. Other parts of children speech, i.e., \textit{test}, \textit{eval\_child\_read} and \textit{eval\_child\_convers}, as well as the adult test speech, i.e., \textit{test\_adult\_1w}, are used for NCM evaluation. \textit{test\_adult\_1w} is part of the original adult test set, which contains $10,000$ utterances. Table \ref{tab:data_child} gives a summary of the above data sets. For each of them, the decoding accuracy produced by the $ASR_{child}$ system is given in terms of character error rate (CER) and sentence error rate (SER). $ASR_{child}$ is an E2E ASR system trained for children speech as described in the next section.



\begin{table}[t]
\centering
\caption{Speech data sets used in the experiments on NCM}
\begin{tabular}{c|c|c|c|c}
\toprule
\multicolumn{2}{c|}{Data set}                & Hours & CER\% & SER\% \\ \hline \hline
Train                 & \textit{dev}                  & 5     & 22.0  & 66.4  \\ \midrule
\multirow{4}{*}{Eval} & \textit{test}                 & 6     & 20.1  & 63.9  \\
                      & \textit{eval\_child\_read}    & 10    & 9.1   & 41.1  \\
                      & \textit{eval\_child\_convers} & 10    & 35.3  & 87.6  \\
                      & \textit{test\_adult\_1w}      & 10    & 26.1  & 84.5  \\ \bottomrule
\end{tabular}
\label{tab:data_child}
\end{table}

\subsection{End-to-end ASR system for children speech}
Utterance-level NCM is evaluated with an E2E ASR system, which is trained to generate Chinese characters from Mandarin speech \cite{ng2020cuhk}. Input features of the ASR system comprise $80$-dimension filter-bank features and $3$ pitch features. Both the shared encoder and the attention decoder adopt the self-attention \textit{Transformer} structure \cite{vaswani2017attention}. There are two different versions of E2E models involved in the following experiments. The system $ASR_{adult}$ is trained only on the adult read speech, i.e., the training set of the CSRC adult speech. The system $ASR_{child}$ is then fine-tuned with children speech in both read and conversational speaking styles, i.e., the training set of the CSRC children speech. 


\subsection{Evaluation metric}
The prediction of NCM is performed as a binary classification task. Thus the Equal Error Rate (EER) and the area under the ROC curve (AUC) are adopted for performance evaluation. These metrics have been commonly used in previous studies on confidence measures. EER refers to the error rate achieved with the operating threshold at which the false acceptance and false rejection rates are equal. AUC measures the average classification performance over the full range of operating threshold. Perfect performance is attained when the EER is equal to 0 or the AUC is equal to 1. 

\subsection{Loss function for NCM module training}
The training labels for NCM module training are determined as follows. If the ASR decoding result on a test utterance perfectly matches with its ground-truth transcription, the label for the NCM classifier is set to 1; otherwise it is set to 0. For the binary classification problem, the binary cross-entropy (BCE) loss is regarded as the default choice. Our preliminary experiment showed that, however, to obtain a stable operating threshold (for determining the EER), which is approximately equal to $0.5$ across different datasets, the weighted focal loss \cite{Lin_2017_ICCV} could be a better choice since it not only handles the class imbalance issue, e.g., SER of $66.4\%$ on \textit{dev} means that $33.6\%$ of the utterances have the label of $1$, but also pays more attention to hard samples. In the present study, the weighted focal loss is adopted as the loss function for NCM module training. 

\section{Results and Analysis}
\subsection{Utterance-level scores}
The results of utterance-level NCM prediction with scores from the top-1 hypothesis are shown as in Table \ref{tab:res1}. The performance with the four component scores described in Section \ref{ssec:predictor_features} are compared. The best performance was achieved by using the attention score, with the EER of $0.2227$ and AUC of $0.8477$.
The two language models' scores are clearly inferior to the CTC and attention scores. This suggests that acoustic information is more pertinent to estimating the confidence score, and linguistic information alone is not sufficient. As shown in Table \ref{tab:res1}, the weighted sum score $\alpha_{weighted}$ which takes direct effect in ASR decoding performs not well (EER: $0.2852$; AUC: $0.7842$) in comparison with $\alpha_{ctc}$ and $\alpha_{att}$. 

\begin{table}[t]
\centering
\caption{NCM Results of set of utterance-level scores of the best hypothesis on the  \textit{test} set. $\alpha_{weighted}$ denotes the weighted sum of the above four component scores. 
($\alpha_{weighted} = 
0.5\alpha_{ctc}+0.5\alpha_{att}+0.8\alpha_{ngram}+0.2\alpha_{rnnlm} $)}
\begin{tabular}{c|cc|c}
\toprule
Predictor feature   & EER    & threshold  & AUC    \\ \midrule  \midrule
$\alpha_{ctc}$      & 0.2445 & 0.4848 & 0.8353 \\ \hline
$\alpha_{att}$      & $\mathbf{0.2227}$ & 0.4848 & $\mathbf{0.8477}$ \\ \hline
$\alpha_{ngram}$    & 0.3847 & 0.5152 & 0.6560 \\ \hline
$\alpha_{rnnlm}$    & 0.4440 & 0.5152 & 0.5777 \\ \hline
$\alpha_{weighted}$ & 0.2852 & 0.5051 & 0.7842 \\ \bottomrule
\end{tabular}
\label{tab:res1}
\vspace{-4mm}
\end{table}

\subsection{NCM based on n-best scores}
\begin{figure}[t]
\centering    \resizebox{\linewidth}{!}{\includegraphics{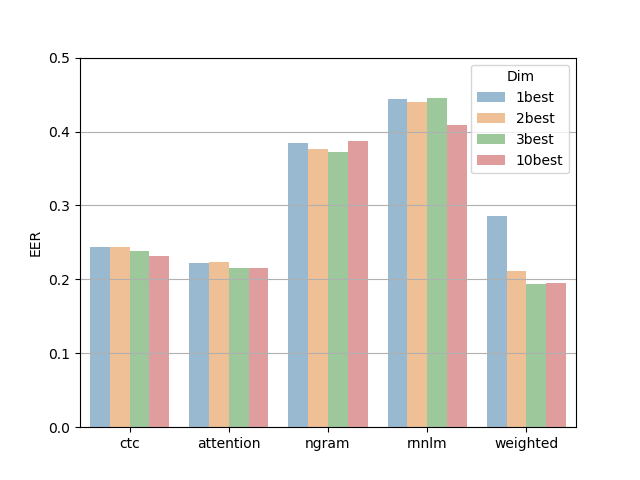}}
    \caption{The performance of $nbest\_score$ features in terms of EER on the \textit{test} set.}
    \label{fig:nbs_eer}
\end{figure}

The experimental results with $nbest\_score$ features are shown by the plots of EER in Figure \ref{fig:nbs_eer}. By incorporating multiple hypotheses ($1best\_score$ to $10best\_score$) into the predictor features, the NCM tends to perform better. Using the $3$best weighted score achieves the best EER of $0.1934$, which significantly surpasses the $1$best weighted score by $0.0918$. It suggests that $nbest\_score$ feature can capture more discriminative pattern to help classification. Particularly, the weighted score outperforms the attention score when it comes to $nbest$ cases. 
The $10best~\alpha_{weighted}$ is considered the most preferred feature since it can achieve similar value of EER ($0.1946$ vs. $0.1934$) to the $3best$ one and higher AUC ($0.8901$ vs. $0.8829$).

\begin{table}[h!]
\caption{NCM Results of $nbest\_score$, $encfeat$, $decfeat$, $avg\_dur$, $top$-$K$ score logits and their variants on the \textit{test} set. The $learn\_weighted$ subscript represents the component weight $\lambda_i$ is jointly learnable via network, the prefix $Ada$ denotes the weight is adaptively adjusted based on the component score. In the rows of $top$-$K$ score logits, $w.~temp$ means the score logits is scaled by a learnable temperature value,  $w.~Adatemp$ means that the temperature value is adaptively adjusted according to the $decfeat$ at each decoder time step via an independent neural network. $prob\_seq$ denotes a sequence of softmax probability  distributions of score logits and $avg~prob~vec$ represents the averaged vector.}

\small
\begin{tabular}{c|c|cc}
\toprule
\multicolumn{2}{c|}{Predictor feature}                                                                                      & EER    & AUC    \\ \midrule \midrule
\multirow{3}{*}{\begin{tabular}[c]{@{}c@{}}1best\\ score\end{tabular}}                   & 
$\alpha_{weighted}$              & 0.2852 & 0.7842 \\   &                                             
$\alpha_{learn\_weighted}$       & 0.2141 & 0.8570 \\
                                                                                         & $Ada$ $\alpha_{learn\_weighted}$ & 0.2152 & 0.8607 \\ \hline
\multirow{3}{*}{\begin{tabular}[c]{@{}c@{}}10best\\ score\end{tabular}}            & $\alpha_{weighted}$       & 0.1946 & 0.8901 \\      
& $\alpha_{learn\_weighted}$       & 0.1929 & 0.8917 \\
                                                                                         & $Ada$ $\alpha_{learn\_weighted}$ & $\mathbf{0.1868}$ & $\mathbf{0.8950}$ \\ \midrule
\multicolumn{2}{c|}{$encfeat$; ~~$dim:~ [T', 256]$}                                                                                              & 0.2647 & 0.8150 \\ \hline
\multicolumn{2}{c|}{$decfeat$; ~~$dim:~ [U, 256]$}                                                                                              & $\mathbf{0.2459}$ & $\mathbf{0.8333}$ \\ \hline
\multicolumn{2}{c|}{$avg\_dur$}                                                                                             & 0.4499 & 0.5668 \\ \midrule
\multirow{3}{*}{\begin{tabular}[c]{@{}c@{}}$top$-$1000$ \\ score \\ logits\end{tabular}} & $ent$                            & 0.3730 & 0.6810 \\
                                                                                         & $ent~w.~temp$                  & 0.2894 & 0.7802 \\
                                                                                         & $ent~w.~Adatemp$               & 0.2300 & 0.8519 \\ \hline
\multirow{3}{*}{\begin{tabular}[c]{@{}c@{}}$top$-$100$\\ score \\ logits\end{tabular}}   & $ent$                            & 0.3097 & 0.7508 \\
                                                                                         & $ent~w.~temp$                  & 0.2803 & 0.7940 \\
                                                                                         & $ent~w.~Adatemp$               & 0.2192 & 0.8616 \\ \hline
\multirow{7}{*}{\begin{tabular}[c]{@{}c@{}}$top$-$20$\\ score \\ logits\end{tabular}}    & $ent$                            & 0.2777 & 0.7941 \\
                                                                                         & $ent~w.~temp$                  & 0.2681 & 0.8093 \\
                                                                                         & $ent~w.~Adatemp$               & 0.2126 & 0.8662 \\ \cline{2-4} 
                                                                                         & $avg$ $prob~vec$                 & 0.2342 & 0.8497 \\
                                                                                         & $avg$ $prob~vec$ $w.~Adatemp$    & 0.2194 & 0.8660 \\ \cline{2-4} 
                                                                                         & $prob\_seq$                      & $\mathbf{0.2089}$ & $\mathbf{0.8771}$ \\
                                                                                         & $prob\_seq$ $w.~Adatemp$         & 0.2097 & 0.8751 \\  \bottomrule
\end{tabular}
\label{tab:res2}
\vspace{-2mm}
\end{table}

\subsection{Other predictor features and variants}
It is worth noting that the component weights $\lambda_i$ can be adjusted automatically. They are treated as learnable parameters jointly optimized in the model training (denoted as $learn\_weighted$ subscript), either in fixed ($\alpha_{learn\_weighted}$) or adaptive way ($Ada~\alpha_{learn\_weighted}$, the prefix $Ada$ denotes the weight is adaptively adjusted based on the component score), as illustrated in the first block of Table \ref{tab:res2}. A large performance gain (EER: $0.2852\xrightarrow{} 0.2152$) can be attained in the case of $1best\_score$, suggesting that balancing the component scores helps a lot. The lowest EER $0.1868$ and the highest AUC $0.8950$ are achieved with the 10best $Ada~\alpha_{learn\_weighted}$ feature.

In the middle part of Table \ref{tab:res2}, the effect of embedding features $encfeat$ and $decfeat$ are compared. The decoder embedding slightly outperforms the encoder one. As a matter of fact, the attention decoder has already attended to the encoder embedding. Yet both of the embeddings are prone to overfit. The $avg\_dur$ is represented by the length ratio ($T'/U$) of $encfeat$ and $decfeat$. It does not provide any discriminative information for confidence measure. 

The lowest part of Table \ref{tab:res2} shows the performance of a few feature variants related to the $top$-$K$ score logits at each time step of decoding for the best hypothesis. The entropy measures the confidence of assigning the maximum index of a softmax probability distribution as the decoding output. That is, the recognition result would be highly uncertain if a distribution is close to a uniform distribution. The average entropy ($ent$) of token distribution is expected to reflect the confidence of an output hypothesis. We experiment with $K = 1000, 100, 20$. It is noted that the entropy with smaller value of $K$, i.e., less $top$-$K$ score logits, consistently outperforms that with larger $K$. Here, $K=20$ is exactly equal to the setting of $beam\_size$. Furthermore, the score logits can be multiplied with a constant temperature value (denoted as $w.~temp$), which is determined by jointly optimizing the NCM module, in order to sharpen or smooth its distribution. Inspired by technique used in \cite{woodward2020confidence}, we can train an additional neural network to predict a dynamic softmax temperature that takes different values at different decoder time steps (denoted as $w.~Adatemp$). Input to this predictive network is the decoder embedding $decfeat$, which contains both acoustic and linguistic information related to the corresponding time step. It can be observed that $ent~w.~Adatemp$ always performs better than $ent~w.~temp$ across different settings of $K$, and both two temperature scaling approaches consistently outperform the vanilla entropy feature. Within the $top$-$20$ score logits, the $ent~w.~Adatemp$ achieves a comparable performance (EER: $0.2126$ vs. $0.2089$; AUC: $0.8662$ vs. $0.8771$) with the $prob\_seq$, though the entropy-based feature is a scalar. 

\subsection{Fusion of predictor features and robustness test}
As shown in the top of Table \ref{tab:res3}, fusion of different predictor features is evaluated on the \textit{test} set of children speech. Generally speaking, fusion of multiple features is beneficial and gives better performance than using them individually. The fusion of $10best~Ada~\alpha_{learn\_weighted}$ and $top$-$20~ent~w.~Adatemp$ features achieve the best NCM quality
(EER: $0.1817$; AUC: $0.9013$), which is exactly the two most potential features observed from the Table \ref{tab:res2}. 

The robustness of NCM is investigated in three different speech domains, namely children read speech, children conversational speech and adult read speech (Table \ref{tab:data_child}). The results are reported as in Table \ref{tab:res3}. It can be observed that in general the performance of NCMs is slightly better on \textit{eval\_child\_convers} set than \textit{test} set, yet make a EER/AUC degradation about $0.2$ on \textit{eval\_child\_read} set.  A large performance gain is attained when being applied on adult speech, which is unexpected. As can be seen, NCM tends to perform better on data sets that have higher sentence error rates (SER). This is probably due to that a high SER (e.g., larger than $65\%$) leads to imbalance distribution of erroneous transcription and correct transcription. This imbalance causes low EER since it becomes easier to separate the two classes.

\begin{table*}[ht]
\centering
\caption{The performance on NCM obtained by fusing predictor features}
\small
\begin{tabular}{c|cc|cc|cc|cc|cc}
\toprule
\multirow{2}{*}{\begin{tabular}[c]{@{}c@{}}Fusion of\\ predictor \\features\end{tabular}} & \multicolumn{2}{c|}{\begin{tabular}[c]{@{}c@{}}$10best~ \alpha_{weighted}$\\ $+~encfeat$\end{tabular}} & \multicolumn{2}{c|}{\begin{tabular}[c]{@{}c@{}}$10best~\alpha_{weighted}$\\ $+~decfeat$\end{tabular}} & \multicolumn{2}{c|}{\begin{tabular}[c]{@{}c@{}}$10best~\alpha_{weighted}$\\ $+~top$-$20~ent$\\ $w.~Adatemp$\end{tabular}} & \multicolumn{2}{c|}{\begin{tabular}[c]{@{}c@{}}$10best~Ada$\\ $\alpha_{learn\_weighted}$\\ $+~top$-$20~ent$\\ $w.~Adatemp$\end{tabular}} & \multicolumn{2}{c}{\begin{tabular}[c]{@{}c@{}}$10best~Ada$\\ $\alpha_{learn\_weighted}$\\ $+~encfeat~+~decfeat$\\ $+top$-$20~ent$\\ $w.~Adatemp$\end{tabular}} \\ \cmidrule{2-11} 
                                                                                        & EER                                                & AUC                                               & EER                                               & AUC                                               & EER                                                         & AUC                                                         & EER                                                                 & AUC                                                                & EER                                                                            & AUC                                                                           \\ \hline \hline
\textit{test}                                                                                  & 0.1846                                             & 0.8980                                            & 0.1841                                            & 0.8939                                            & 0.1811                                                      & 0.9003                                                      & $\mathbf{0.1817}$                                                              & $\mathbf{0.9013}$                                                             & 0.1830                                                                         & 0.8949                                                                        \\
\textit{eval\_child\_read}                                                                       & 0.2083                                             & 0.8777                                            & 0.2097                                            & 0.8753                                            & 0.2034                                                      & 0.8797                                                      & $\mathbf{0.2034}$                                                              & $\mathbf{0.8813}$                                                             & 0.2091                                                                         & 0.8736                                                                        \\
\textit{eval\_child\_convers}                                                                    & 0.1788                                             & 0.9021                                            & 0.1755                                            & 0.9005                                            & 0.1788                                                      & 0.9039                                                     & $\mathbf{0.1748}$                                                              & $\mathbf{0.9056}$                                                             & 0.1765                                                                         & 0.9047                                                                        \\
\textit{test\_adult\_1w}                                                                        & 0.1641                                             & 0.9193                                            & 0.1607                                            & 0.9174                                            & 0.1712                                                      & 0.9152                                                      & $\mathbf{0.1538}$                                                              & $\mathbf{0.9231}$                                                             & 0.1748                                                                         & 0.9106                                                                        \\ \bottomrule
\end{tabular}
\label{tab:res3}
\end{table*}

 \begin{table}[h]
 \centering
 \caption{Performance comparison of NCM on two different ASR domains. The NCM matches with $ASR_{child}$.}
 \small
\begin{tabular}{c|c||c|c}
\toprule
\multicolumn{2}{c|}{Evaluation set}         & \begin{tabular}[c]{@{}l@{}}decoded by\\ $ASR_{child}$\end{tabular} & \begin{tabular}[c]{@{}l@{}}decoded by\\ $ASR_{adult}$\end{tabular} \\ \midrule
\multirow{2}{*}{\textit{test}}                 & EER & 0.1946                                                            & $\mathbf{0.1621}$                                                            \\
                                      & AUC & 0.8901                                                            & $\mathbf{0.9176}$                                                            \\ \hline
\multirow{2}{*}{\textit{eval\_child\_read}}    & EER & $\mathbf{0.2039}$                                                            & 0.2272                                                            \\
                                      & AUC & $\mathbf{0.8809}$                                                            & 0.8602                                                            \\ \hline
\multirow{2}{*}{\textit{eval\_child\_convers}} & EER & $\mathbf{0.1829}$                                                            & 0.1867                                                            \\
                                      & AUC & $\mathbf{0.8981}$                                                            & 0.8862                                                            \\ \hline
\multirow{2}{*}{\textit{test\_adult\_1w}}      & EER & $\mathbf{0.1662}$                                                            & 0.2153                                                            \\
                                      & AUC & $\mathbf{0.9195}$                                                            & 0.8654                                                            \\ \bottomrule
\end{tabular}
\label{tab:res4}
\vspace{-2mm}
\end{table}

\subsection{Confidence measure on mismatched ASR}
Different ASR systems may exhibit different decoding behaviours. A confidence measure module is designed typically for a specific ASR system. The transferability of our proposed NCM module, i.e., how well it can be used with another ASR system, is investigated in this section. The $ASR_{adult}$ system is taken as a mismatched ASR system against $ASR_{child}$. The performance of the NCM with feature $10best~ \alpha_{weighted}$ is shown as in Table \ref{tab:res4}. A clear EER improvement from $0.1946$ to $0.1621$ is observed on the \textit{test} set. And the EER performance degraded from $0.1662$ to $0.2153$ on the \textit{test\_adult\_1w} set. Decoding children speech utterances by $ASR_{adult}$ would produce more erroneous transcriptions (higher SER), resulting in more imbalanced class distribution. In this case, EER/AUC seems not to be appropriate for performance comparison. 
 
\begin{figure}[t]
    \centering
    \resizebox{\linewidth}{!}{\includegraphics{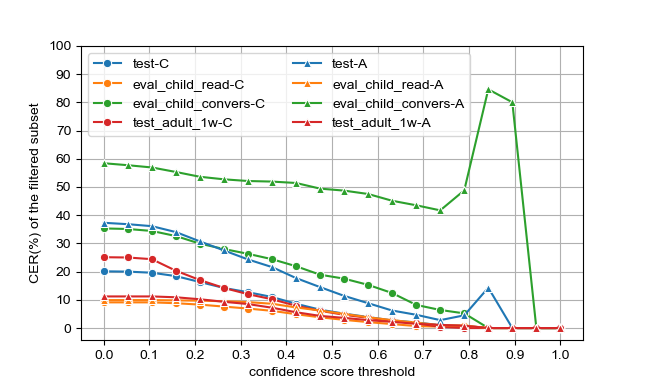}}
    \caption{CERs of filtered utterances w.r.t confidence threshold for different evaluation sets that decoded by $ASR_{child}$ (denoted by suffix \textit{C} with circle marker) and $ASR_{adult}$ (denoted by suffix \textit{A} with triangle marker), respectively. }
    \label{fig:cer}
\end{figure}

Motivated by the work in \cite{li2020confidence}, we evaluate the performance of NCM by plotting the CER on filtered utterances with respect to the confidence threshold. Utterances with confidence score higher than a specific threshold are selected to form a set of filtered utterances. Since a good NCM should exhibit a strong correlation with the CER, i.e., a higher threshold will result in a set of utterances with lower CER, a monotonically decreasing relation is expected. As shown in Figure \ref{fig:cer}, most curves show a trend of monotonic descending. Nevertheless, two prominent spikes are noted in the region of high confidence for the children speech in the \textit{test} and \textit{eval\_child\_convers} data sets decoded by $ASR_{adult}$. The spikes reveal the over-confidence behaviour related to transferability. We suspect this transferability issue is highly related to decoding the children conversational speech since it is the common speech data type in  both \textit{test} and \textit{eval\_child\_convers} sets.
 

\section{Conclusions}
This research is focused on investigating the efficacy of NCMs which are derived from different predictor features in E2E ASR systems. It is found that properly balanced weights on the CTC score, attention score and the language model scores play a critical role in the reliability of confidence measure. Incorporating the n-best hypothesis scores can lead to further improvement. In addition, the average token entropy with adaptive softmax temperature is demonstrated to be effective.
The fusion of these features can achieve better performance. 
Experimental results also suggest that the EER/AUC metrics are not sufficient to evaluate the NCM performance on a mismatched ASR with large SER difference.



\bibliographystyle{IEEEbib}
\bibliography{refs}

\end{document}